\documentclass[useAMS,usenatbib]{mn2e}

\usepackage{graphicx}
\usepackage{amssymb}

\title[The red sequence of AXU clusters]{The red 
sequence of Abell X-ray underluminous clusters}
\author[Trejo-Alonso et al.]{J. J. Trejo-Alonso$^{1,}$$^{2}$
\thanks{E-mail:josue@astro.ugto.mx},
C. A. Caretta$^{1}$, 
T. F. Lagan\'a$^{2,}$$^{3}$, 
L. Sodr\'e Jr.$^{2}$, 
\and E. S. Cypriano$^{2}$, 
G. B. Lima Neto$^{2}$,
C. Mendes de Oliveira$^{2}$
\\
$^{1}$ Departamento de Astronom\'ia, Universidad de Guanajuato, 
Callej\'on de Jalisco S/N, Valenciana, C.P.: 36240, Guanajuato, 
Gto., M\'exico.\\
$^{2}$ Instituto de Astronomia, Geof\'isica e Ci\^encias 
Atmosf\'ericas, Universidade de S\~ao Paulo, Rua do Mat\~ao 1226, 
Cidade Universit\'aria, \\CEP: 05508-090, S\~ao Paulo, SP, Brasil.\\
$^{3}$ N\'ucleo de Astrof\'isica Te\'orica, Universidade Cruzeiro 
do Sul, Rua Galv\~ao Bueno 686, Liberdade, CEP:
01506-000, S\~ao Paulo,\\ SP, Brasil.\\
}
\begin{document}

\date{Accepted 2014 March 25. Received 2014 March 24; in original form 2013 June 22}

\pagerange{\pageref{firstpage}--\pageref{lastpage}} \pubyear{2013}

\maketitle

\label{firstpage}

\begin{abstract}
We present an analysis of the colour-magnitude relation for 
a sample of 56 X-ray underluminous Abell clusters, 
aiming to unveil properties that may elucidate the evolutionary 
stages of the galaxy populations that compose such systems. 
To do so, we compared the parameters of their colour-magnitude 
relations with the ones found for another sample of 50 
``normal'' X-ray emitting Abell clusters, both selected in 
an objective way. The $g$ and $r$ magnitudes from the SDSS-DR7 were 
used for constructing the colour-magnitude relations. 
We found that both samples show the same trend: the red sequence slopes 
change with redshift, but the slopes for X-ray underluminous clusters 
are always flatter than those for the normal clusters, by a difference of 
about 69\% along the surveyed redshift range of 0.05 $\le z <$ 0.20. 
Also, the intrinsic scatter of the colour-magnitude relation 
was found to grow with redshift for both samples but, for the 
X-ray underluminous clusters, this is systematically larger by about 
28\%. By applying the Cram\'er test to the result of this
comparison between X-ray normal and underluminous cluster samples, 
we get probabilities of 92\% and 99\% that the red sequence slope 
and intrinsic scatter distributions, respectively, differ, in the 
sense that X-ray underluminous clusters red sequences show flatter 
slopes and higher scatters in their relations. 
No significant differences in the distributions of red-sequence 
median colours are found between the two cluster samples. 
This points to X-ray underluminous clusters being younger systems than 
normal clusters, possibly in the process of accreting 
groups of galaxies, individual galaxies and gas.
\end{abstract}


\section{Introduction}

Since \cite{baum} noticed that, within a sample of elliptical galaxies, 
the brightest are generally the reddest ones, 
several studies have demostrated that the red galaxy 
population traces a straight line in colour-magnitude diagrams, with a well-defined 
slope and small scatter  \citep[$\le$ 0.1 mag; e.g.][]{Vis77,Bower3,vandokkum98,
andreon,lopez,mcintosh}. This makes the colour-magnitude relation (CMR) a 
useful tool for getting information about the formation and evolution of galaxy 
cluster members.

The CMR can be used for estimating the redshift of a cluster when 
photometry in at least two bands is available \citep[e.g.][]{Vis77, 
Bower3, andreon, lopez},
while a blind search for CMRs may be used for finding galaxy overdensities 
associated to galaxy clusters \citep[e.g.][]{Gladders00}.
The slope of the CMR has been interpreted as a consequence of the mass-metallicity
relation: more massive (or luminous) galaxies have redder colours 
\citep[e.g.][]{Faber, arimoto, kodama97}.
The mass-metallicity relation, on the other hand, is thought to be originated 
from the tendency of galaxies to lose their metals due to galactic 
winds, the loss being more pronounced for galaxies with 
shallower potential wells or lower masses 
\citep[e.g.][]{arimoto,Kodama,vandokkum01,Tremonti,Gallazzi}.
An alternative model \citep{kauffmann98} considers the effect of the formation
of elliptical galaxies by the merging of disk systems: the mass-metallicity
relation would be already established for the progenitors, and the 
more massive ellipticals are formed from more massive disks.
Other works have shown that the CMR depends neither on cluster/group richness 
\citep{andreon} nor on the environment \citep{Hogg}, but luminous
galaxies are more abundant in higher over-density regions and blue galaxies 
reside towards the outskirts of the clusters \citep{mcintosh,Hogg}, 
a reminiscence of the morphology-density relation \citep[e.g.][]{dressler80}.

The physical interpretation of these relations and properties is not fully 
understood. The CMR tight scatter has been interpreted as evidence that galaxies 
residing in this sequence are coeval, with a small spread in their formation age, 
which mostly occurred at $z\,\ge\,$2.0 and evolved passively since then 
\citep[e.g.][]{Bower3,Bower,Bower4,Lubin3,ellis,Gladders,Kodama,vandokkum98,
andreon,mcintosh}. Other works have suggested that the age of the galaxies could 
drive the CMR \citep[e.g.][]{Ferreras,Terlevich,trager00}. Some studies, like 
\citet{Skelton} and \citet{jimenez11}, have considered the effect of gas-poor mergers 
on the CMR with a toy model, concluding that the relation changes its slope at the 
bright end and becomes bluer and with lower scatter. It is worth mentioning that, 
besides ellipticals, lenticulars and passive spirals also populate the CMR
\citep[e.g.][]{bamford09,masters10,sodre13}.

On the other hand, the majority of galaxy clusters have most of
their baryons in the form of a hot, diffuse plasma, that may interact
with the colder interstellar gas (interstellar medium) due mainly to the
motion of galaxies inside clusters. 
The ram-pressure effect \citep{gunn72,larson80}, i.e. the hydrodynamical 
interaction between the intracluster medium (ICM) and the interstellar medium,
may strip the gas of the galaxies, quenching star formation or 
even enhancing it while the gas is being compressed, affecting galaxy 
evolution and hence the colours of cluster galaxies \citep{fujita99,weinberg13}.

The ICM properties, such as luminosity and temperature, are observed to 
scale with the mass of the cluster, but possibly not in a self-similar 
manner, as described by the scaling relations between X-ray luminosity 
or temperature and virial mass \citep[$L_\mathrm{X}$-$M_\mathrm{v}$ 
and $T_\mathrm{X}$-$M_\mathrm{v}$, e.g.][]{Reiprich,Rykoff}, and X-ray 
temperature with galaxy velocity dispersion \citep[$T_\mathrm{X}$-$\sigma$, 
e.g.][]{Lubin,Xue}.

In recent years, several studies have found galaxy systems with 
most properties of galaxy clusters but underluminous in X-rays 
\citep{Bower,Lubin2,Dietrich}, with respect to the scaling relations 
mentioned above. Dynamical analyses of these systems suggest that they are 
young systems still undergoing a phase of gas accretion and/or merging 
of smaller groups and galaxies; that is, they may not have yet 
reached the virial equilibrium. 
Other works, however, debate the very existence of these objects 
\citep[e.g.,][]{andreon2}.

Our objective with the present work is to investigate the properties 
of the CMR of X-ray Underluminous Abell clusters (AXUs, following 
\citealp{p1} naming) in comparison with ``Normal'' X-ray emitting Abell 
clusters (AXNs) to verify whether the ICM is indeed important in 
shaping up the red sequence. 
The paper is organized as follows: the optical and X-ray data are 
described in \S\ref{data}. The identification of the red sequence, 
its analysis and comparisons between samples are presented in 
\S\ref{analysis}. Our main results are summarized in \S\ref{results} 
and final comments and conclusions in  \S\ref{discussion} and 
\S\ref{conclusion}, respectively.

We will assume a $\Lambda$CDM cosmology with $\Omega_m=$ 0.30, 
$\Omega_\Lambda=$ 0.70 and $H_0=$ 70 km s$^{-1}$ Mpc$^{-1}$ throughout 
the paper.

\section{Data}
\label{data}
The analysis presented in this paper is based on a comparison of the CMR 
properties between a sample of AXUs and a sample of AXNs.
The colour data for both samples were extracted 
from the Sloan Digitized Sky Survey, Data Release 7 \citep[SDSS-DR7,][]{a1}.

\begin{figure}
\includegraphics[width=\linewidth]{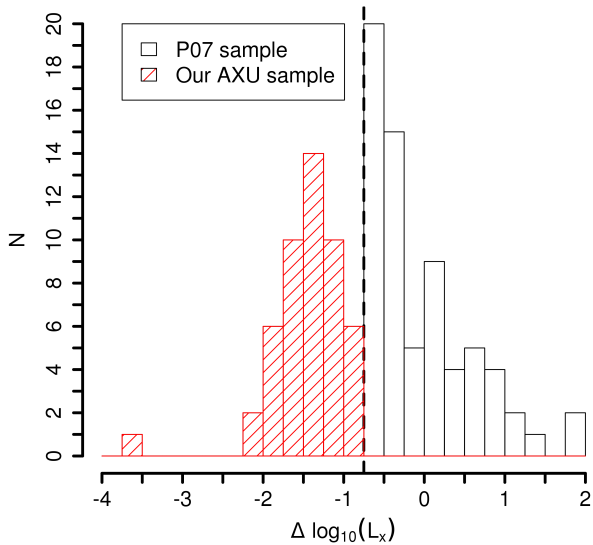}
\caption{Distribution of the residuals for P07 sample, defined as the 
difference between measured and predicted (by the scaling relation for 
AXNs) cluster 
X-ray luminosity ($\Delta\log_{10}{\mathrm{L_X}}\,=\,\log_{10}\mathrm{L_{X,m}}-
\log_{10}\mathrm{L_{X,p}}$). All clusters below $\Delta\log_{10}
{\mathrm{L_X}}\le-0.75$  
(hatched histogram) are considered as our AXU sample.}
\label{histdeltalx}
\end{figure}

The sample of AXUs comes from that of the optically selected clusters 
by \citet[][hereafter P07]{p1}; their original sample was defined from the 
\citet{abell58} cluster catalog in the region covered by Sloan Digitized 
Sky Survey, Data Release 3 (SDSS-DR3, \citealt{y1}).
That sample contains 137 spectroscopically confirmed clusters, 
51 of which were only marginally detected (between $2$ and $3\sigma$)
in the ROSAT All Sky Survey data (RASS, \citealt{v1})
or presenting no clear X-ray emission (detection level $\sim\,1\sigma$). 
The authors determined X-ray luminosities and virial masses in order to 
analyze the $L_\mathrm{X}$-$M_{200}$ relation. 
They classified these 51 clusters as AXU clusters based on the
poorly detected or undetected signal of X-ray emission and the fact 
that, on average, these did not follow the $L_\mathrm{X}$-$M_{200}$ 
scaling relation. However, since the first criterion
is subject to debate, because it is not clear if significant X-ray emission
may be necessarily expected for all the clusters in the sample (mainly
due to the shallow depth of the RASS), and the second is 
only a statistical one, we applied an independent criterion to
separate the AXU clusters. We analyzed the $L_\mathrm{X}$-$M_{200}$ 
relation for the optically selected clusters of P07 and 
defined as AXUs the clusters (56 systems) with 
$\Delta\log{\mathrm{L_X}} \le -0.75$, 
where the $\Delta\log{\mathrm{L_X}}$ is the difference between the 
calculated and predicted cluster X-ray luminosity (Figure \ref{histdeltalx}), 
according to the same scaling relation presented in P07.
There is no clear argument for using this threshold for separating 
the AXU clusters. Since this sample contains both AXUs and AXNs, we only 
require that the AXUs have the most negative values of 
$\Delta\log{\mathrm{L_X}}$. However, the P07 scaling relation was obtained
for AXN clusters only (in fact the same sample we use in the present
paper, to be described below) and the original fit has a standard deviation
of 0.43 in $\Delta\log{\mathrm{L_X}}$, which means that our separating line
is roughly about 2$\sigma$ from the scaling relation (that is, 92\% 
of the AXNs should be excluded by applying this threshold), making us confident 
that we are close to selecting a bona fide AXU's population.

\begin{figure}
\includegraphics[width=\linewidth]{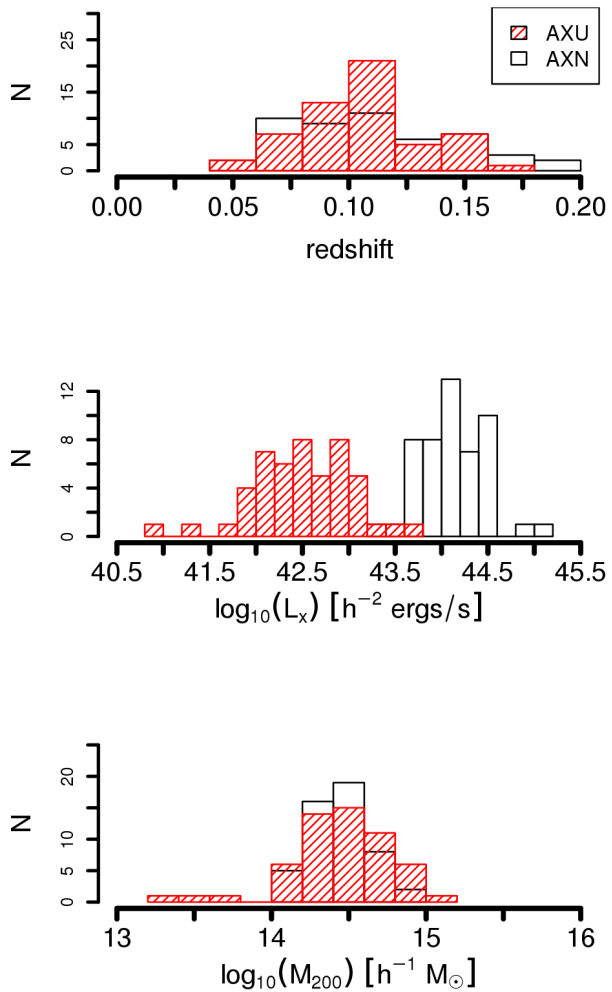}
\caption{General properties for the sample of 56 AXUs (hatched histograms) and 
50 AXNs (solid line histograms).}
\label{propsam}
\end{figure}

In addition, P07 also present an estimate of the number of cluster members 
(within 1 Abell radius), redshift ($z$), velocity dispersion ($\sigma_v$),
cluster virial radius ($r_{200}$), \citet{d1} probability that the cluster 
does not contain substructures (P$_{\mathrm{DS}}$), and an X-ray class 
(a classification based on the quality of the X-ray detection). For such AXU sample, 
we extracted position and photometric data (in the $g$ and $r$ bands, 
corrected for extinction from GALAXY and DERED 
tables, \citealt{Strauss}) from SDSS-DR7 for galaxies brighter than 
$r\,=\,22$ within a projected $r_{200}$ radius of each cluster centre 
(taken directly from P07). Figure \ref{propsam} shows some general 
properties of the AXU sample (shaded histograms). 
The median (mean) values of some properties of these clusters are presented 
in Table \ref{propsamples}.

\begin{table*}
  \centering
\begin{tabular}{ccrrr}
\hline \hline
\multicolumn{5}{c}{Table \ref{propsamples}. General properties: median 
(mean) values}\\ 
\hline
\multicolumn{1}{c}{Sample} & 
\multicolumn{1}{c}{$r_{200}$} &  
\multicolumn{1}{c}{$z$} &
\multicolumn{1}{c}{$M_{200}$} &
\multicolumn{1}{c}{$L_\mathrm{X}$} \\
\multicolumn{1}{c}{} & 
\multicolumn{1}{c}{$h^{-1}$Mpc} &  
\multicolumn{1}{c}{} &
\multicolumn{1}{c}{10$^{14}\,h^{-1}$M$_\odot$} &
\multicolumn{1}{c}{10$^{44}\,h^{-2}$ergs s$^{-1}$} \\
\hline \hline

AXU & 1.08 (1.10) & 0.106 (0.106) & 2.90 (3.59) & 0.03 (0.05) \\
AXN & 0.98 (1.00) & 0.113 (0.111) & 2.67 (3.02) & 1.29 (1.90) \\

\hline \hline
\end{tabular}
\caption{General properties of the AXN and AXU samples. Note that, 
although the median (mean) mass of AXU clusters is slightly higher, 
their luminosties are significantly lower.}
\label{propsamples}
\end{table*}

Our comparison sample comes from the list of 114 ``normal'' X-ray emitting 
clusters described in \citet[][hereafter P04]{p2}. This sample was analyzed 
in the same way as the AXU sample and has 91 galaxy clusters with $M_{200}$ 
and $r_{200}$ information from \citet{piffaretti11}, but only 50 are 
in the same redshift range of our AXU sample, and they will be our AXN sample. 
Figure \ref{propsam} also shows the properties of the AXN sample (solid 
line histograms). The median (mean) values of these properties are also 
presented in Table \ref{propsamples}.

\begin{figure}
\includegraphics[width=\linewidth]{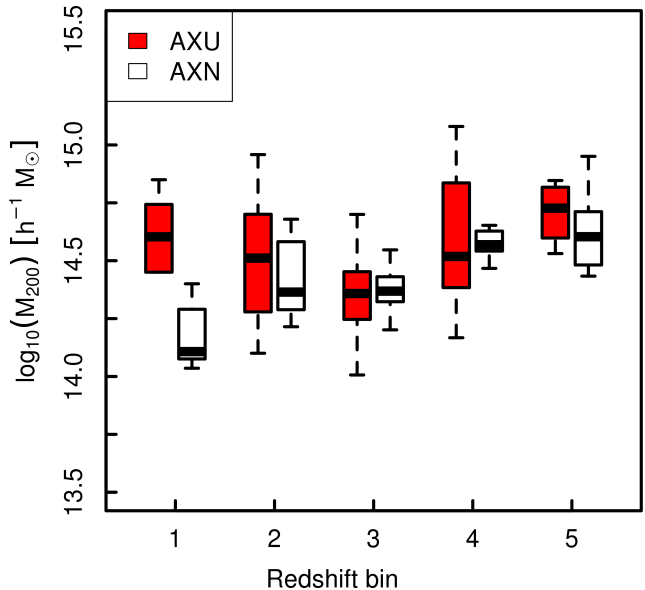}
\caption{Boxplots of the median mass comparison between AXUs and AXNs
in the different redshift bins considered for CMR analyses. The filled boxes 
(red in the electronic version) are the AXU cluster mass range and the white 
ones are the AXNs. The black thick lines inside boxes are the median mass 
values while 
the large boxes are the inter-quartile ranges (defined as the
difference between the third and first quartiles) of the mass distribution. 
Also, the  ``error bars'' are the first and 
third quartile minus 1.5 times the inter-quartile range.}
\label{massbinzbox}
\end{figure}

From now on, we will divide the samples in 5 redshift bins (details below). 
In Figure \ref{massbinzbox} we test if our division in bins is sufficiently 
homogeneous for properly comparing the red sequence fits. We can see that 
the median mass is almost the same for all bins (0.075$\,<\,z\,<\,$0.200), 
except for the first bin, being slightly displaced in the sense that the 
clusters in our AXU sample are in general more massive than AXNs. 
We believe this does not affect significantly our results since most of the 
signal is in the three intermediate bins. On the other hand, this may point 
out that AXU clusters, being possibly younger systems, are not necessarily 
less massive (they may still be currently accreting clumps of galaxies and 
increasing their masses) or could have their masses overestimated due to 
their unrelaxed dynamical state.

\section{Analysis}
\label{analysis}
A rest-frame analysis of the CMR of a sample covering a range in 
redshifts requires k-correction of galaxy magnitudes. 
Since the k-correction depends also on the SED of the galaxy 
(usually approximated, for practical purposes, by using a colour index 
or morphological class), some artificially large spread may be expected, 
especially at higher redshifts, due to the uncertainties in the SED 
proxy parameters. Thus, despite the fact that there are recent useful 
k-correction estimations for SDSS-DR7 galaxies 
\citep[e.g.][]{omill11}, we decided to adopt a more robust approach, by 
comparing the CMR parameters in different redshift bins, with no k-corrections 
applied. 

We divided our samples in five sub-samples comprising clusters within bins 
of 0.025 in redshift (except for the farthest one): 0.05$\,\leq\,z\,<\,$0.075, 
0.075$\,\leq\,z\,<\,$0.10, 0.10$\,\leq\,z\,<\,$0.125, 0.125$\,\leq\,z\,<\,$0.15 and 
0.15$\,\leq\,z\,<\,$0.20. The cluster galaxies (without k-correction),
in each redshift bin, were stacked in the corresponding sub-sample. We, then,
constructed density plots using a Gaussian kernel taking the 
peak value of the distribution as the centre of the red sequence 
(we also tested Epanechnikov, rectangular and biweight kernels, 
with the same peak value as result) and the bandwidth as the standard 
deviation of the kernel\footnote{All the 
computations in this paper were done with the {\sf R} software 
http://www.r-project.org/ \citep{R}.}. 
Then, the limits for the colour range inside which the CMR was
fitted were defined as $\pm$0.30 from the peak value.
For the luminosity intervals, we have considered as the brigthest 
limit one magnitude brighter than the mean $r$ magnitude of the BCGs 
in the MaxBCG catalogue (\citealt{k1}) in the same redshift bin, 
$\left \langle r_{BCG}\right \rangle -1$, and as the faintest limit 
$\left \langle\,r_{BCG} \right \rangle +4$, a reliable limit taking 
into account the SDSS completeness \citep[e.g.][]{yasuda01}. 
Although the limits adopted for the present work were those just described (listed 
in Table \ref{tabla1} for each redshift bin), we also made two other 
tests: for the first one, 
we considered as the upper limit in $r$ magnitude $\left \langle 
r_{BCG}\right \rangle +5$ and, for the second one, we took 
as the lower limit in $r$ magnitude one magnitude brighter 
than the mean $\left \langle M_r^*\right \rangle$, using the 
Schechter parameters from  \cite{popesso05}, in the same redshift 
bin, converted to apparent magnitude, and as the upper limit 
$\left \langle M_r^*\right \rangle +4$, obtaining, for both alternative
selections, similar results. 

\begin{table*}
  \centering
\begin{tabular}{ccrr}
\hline \hline
\multicolumn{4}{c}{Table \ref{tabla1}. Adoppted limits for red sequence fit per redshift bin}\\ \hline
\multicolumn{1}{c}{Redshift bin} & 
\multicolumn{1}{c}{Magnitude range} &  
\multicolumn{1}{c}{Galaxies in AXUs} &
\multicolumn{1}{c}{Galaxies in AXNs}\\ 
\hline \hline

0.050$\,\leq\,z\,<\,$0.075 & 13.52-18.52 & 27865 & 18029\\ 
0.075$\,\leq\,z\,<\,$0.100 & 14.28-19.28 & 41449 & 24441\\ 
0.100$\,\leq\,z\,<\,$0.125 & 14.99-19.99 & 24963 & 10202\\ 
0.125$\,\leq\,z\,<\,$0.150 & 15.56-20.56 & 10028 & 7612\\ 
0.150$\,\leq\,z\,<\,$0.200 & 16.19-21.19 &  4552 & 4919\\ 
\hline \hline
\end{tabular}
\caption[Summary of the used sub-samples]{Luminosity limits for the selection of 
red sequence galaxies in bins of redshift for the AXU and AXN samples. 
The lower and upper magnitude limits are $\left \langle r_{BCG}\right 
\rangle -1$ and $\left \langle r_{BCG} \right \rangle +4$, respectively. 
We show the number of galaxies, in the field, used for the analysis 
in each sub-sample.}
\label{tabla1}
\end{table*}

\begin{figure}
\includegraphics[width=\linewidth]{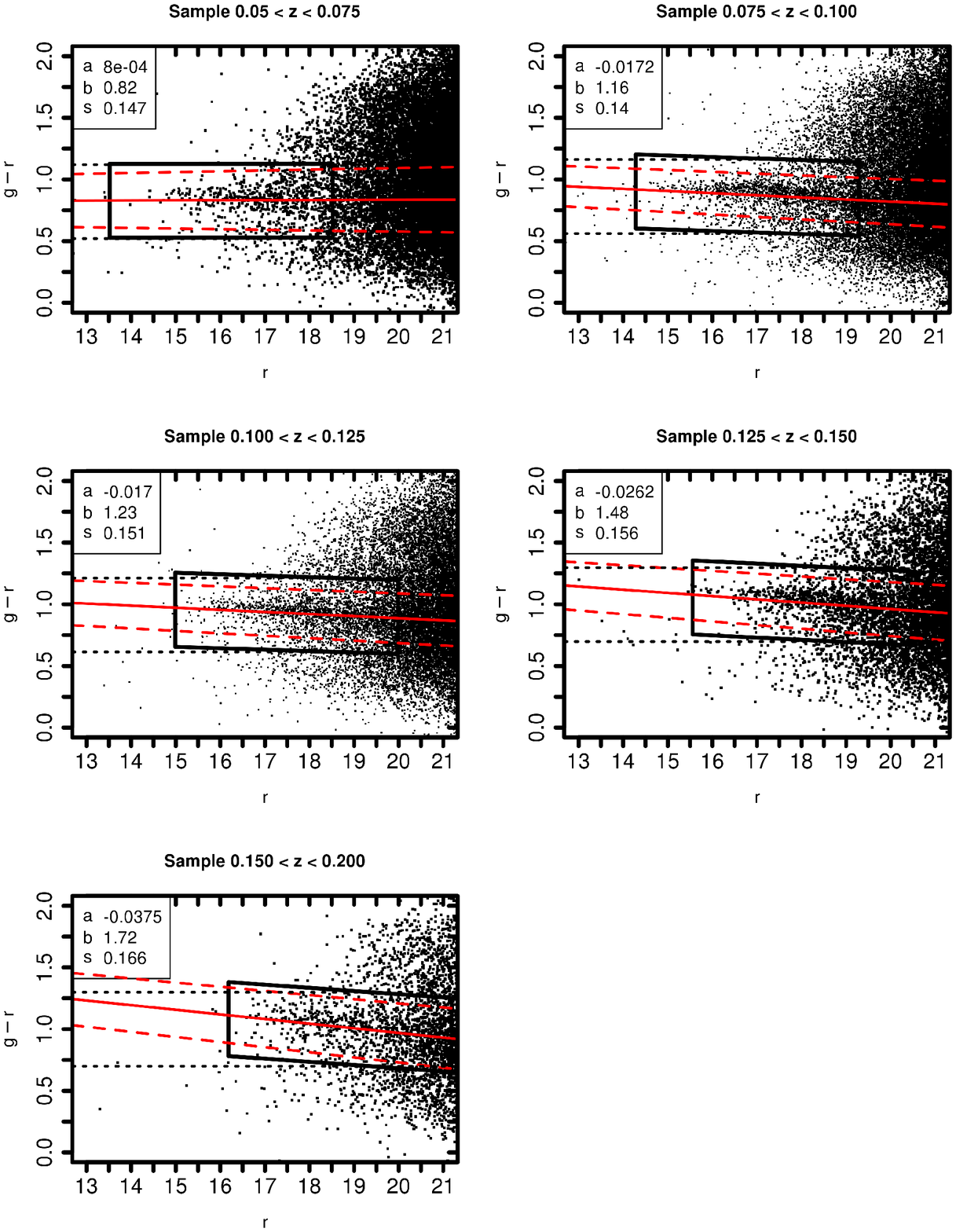}
\caption{Colour-magnitude diagrams for each sub-sample of the AXU sample.  The dotted lines are the first selection in colour and the solid rectangles (black in the electronic version) are the final red sequences selections. The solid lines are the regression fit and the dashed lines are 1$\sigma$ errors from fit (both red in the electronic version).}
\label{cmrzbin2}
\end{figure}

\begin{figure}
\includegraphics[width=\linewidth]{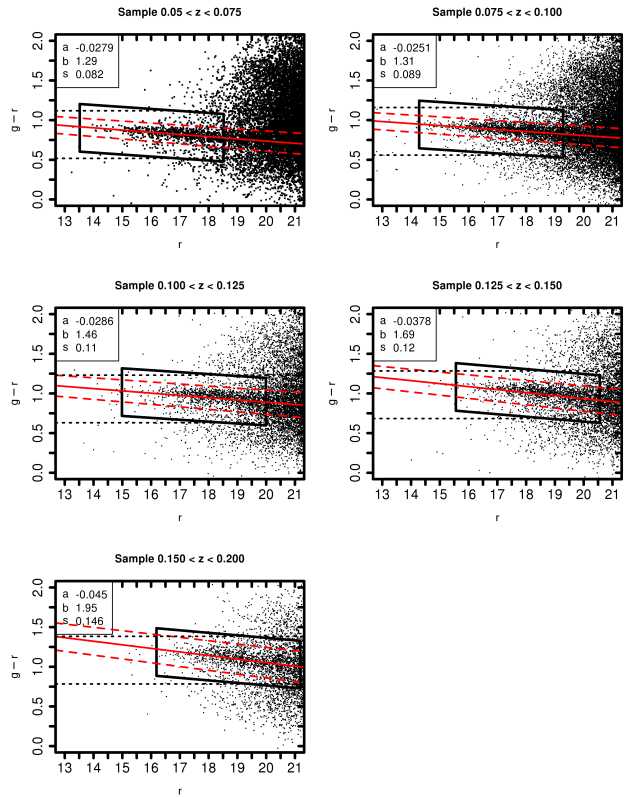}
  \caption{Colour-magnitude diagrams for each bin of redshift in the AXN sample.}
\label{cmrzbinpop042}
\end{figure}

For retrieving the slope and intercept of the stacked CMRs, after 
the first definition of limits in colour and luminosity, we have made 
a robust regression, using the MM-estimator included in the MASS 
package (\citealt{v2}) in {\sf R} (fitting is done by iterated re-weighted 
least squares, IWLS, where standard deviation of the error term is 
not constant over all values of the predictor or explanatory variables). 
Then, we redefined the boundaries of the box taking lines parallel 
to the previous fit, with the upper and lower limits on $\pm$0.30. 
The same fitting procedure was then repeated. 
This selection is similar to other recent choices in the 
literature (e.g. \citealt{d2,l1}), but our results are not 
qualitatively sensitive to this particular choice of $\pm$0.30, 
as we will see later. From the new fit we obtained the slope, intercept, 
intrinsic scatter, median and mean colours (Figures \ref{cmrzbin2} and 
\ref{cmrzbinpop042}). The solid line, in both figures, is the 
fit by the robust regression (MM-estimator) of the red sequence 
while the dashed lines are 1$\sigma$ error bars from the fit. 
The values in the boxes are: slope (a), intercept (b) and 
intrinsic scatter (s).

\begin{table*}\small
\begin{center}
\begin{minipage}{180mm}
\begin{tabular}{cccccccccc}
\hline \hline 
\multicolumn{10}{c}{Table \ref{tabla2}. Parameters of the red sequences}\\ \hline
 &  \multicolumn{4}{c}{AXU} & & \multicolumn{4}{c}{AXN} \\
 \cline{2-5}  \cline{7-10}\\
Redshift bin & Slope & Intercept & Scatter &
 Colour & & Slope& Intercept & Scatter & 
  Colour \\ 
\hline \hline

0.050$\,\leq\,z\,<\,$0.075 &    0.0008$\pm$0.0058 & 0.08$\pm$0.10 & 0.147 & 0.841$\pm$0.019 & & $-$0.0279$\pm$0.0029 & 1.29$\pm$0.05 & 0.082 & 0.822$\pm$0.037\\
0.075$\,\leq\,z\,<\,$0.100 & $-$0.0172$\pm$0.0027 & 1.16$\pm$0.05 & 0.140 & 0.867$\pm$0.019 & & $-$0.0251$\pm$0.0019 & 1.31$\pm$0.03 & 0.089 & 0.865$\pm$0.012\\
0.100$\,\leq\,z\,<\,$0.125 & $-$0.0170$\pm$0.0026 & 1.23$\pm$0.05 & 0.151 & 0.917$\pm$0.021 & & $-$0.0286$\pm$0.0027 & 1.46$\pm$0.05 & 0.109 & 0.938$\pm$0.015\\
0.125$\,\leq\,z\,<\,$0.150 & $-$0.0262$\pm$0.0034 & 1.48$\pm$0.07 & 0.156 & 0.989$\pm$0.021 & & $-$0.0378$\pm$0.0026 & 1.69$\pm$0.05 & 0.120 & 0.983$\pm$0.016\\
0.150$\,\leq\,z\,<\,$0.200 & $-$0.0375$\pm$0.0039 & 1.72$\pm$0.08 & 0.170 & 0.981$\pm$0.024 & & $-$0.0450$\pm$0.0029 & 1.95$\pm$0.06 & 0.146 & 1.075$\pm$0.020\\

\hline \hline
\end{tabular}
\caption{{\small Parameters obtained after fitting the red sequences. 
For each sub-sample we get the slope, intercept, intrinsic scatter and 
median colour.}}
\label{tabla2}
\end{minipage}
\end{center}
\end{table*}

\section{Results} 
\label{results}

\begin{figure}
\includegraphics[width=\linewidth]{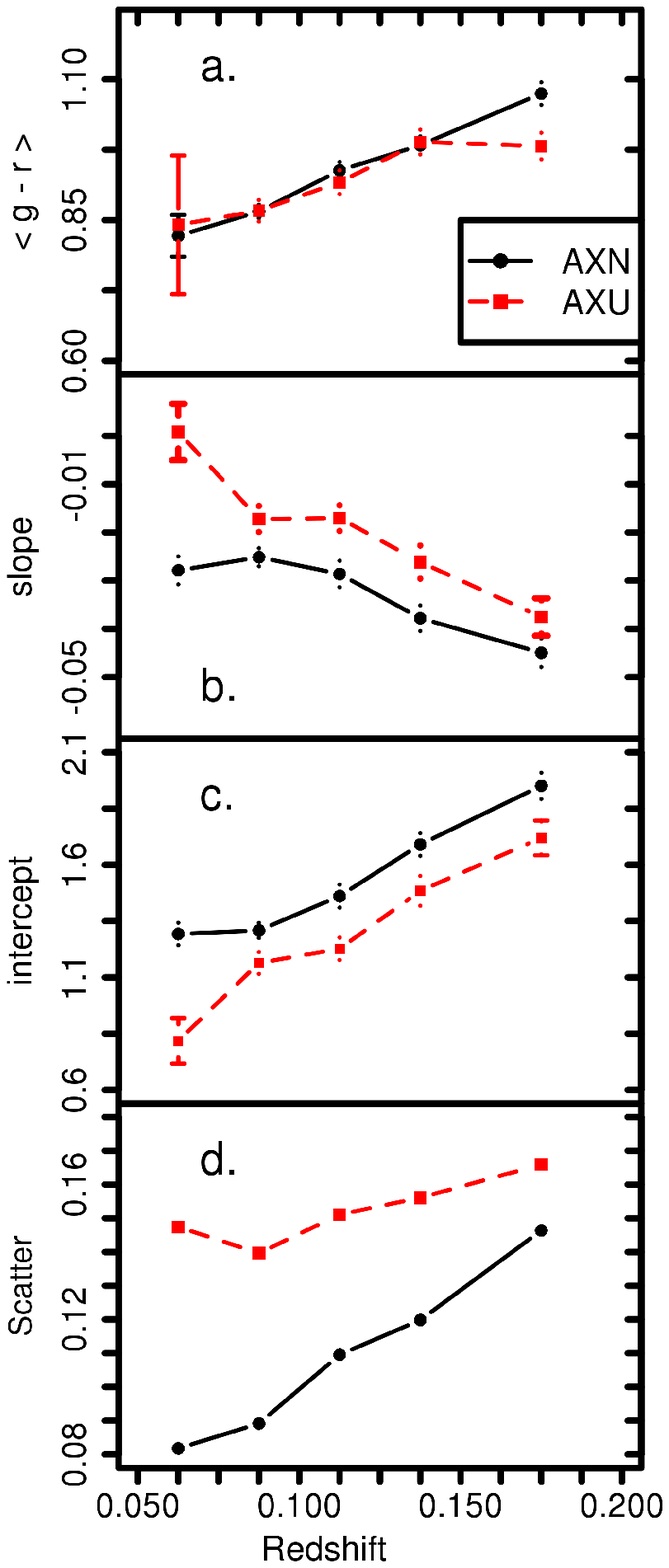}
\caption{Results of the red sequence fitting for the two samples: AXN 
(closed squares and solid lines) and AXU (closed circles and dashed lines). 
Errors were obtained with the MM-estimator, except for the error of the median, 
which was obtained by bootstraping.}
\label{statistics2}
\end{figure}

The results of the analysis explained in the previous section are presented 
in Table \ref{tabla2}. Figure \ref{statistics2} shows that there are clear 
trends of CMR properties with redshift, in the sense that intercept, 
intrinsic scatter and median colour decrease with decreasing redshift. 
The slope seems to flatten with decreasing redshift.

The most obvious trend is the increase of the median colour with redshift 
(Figure \ref{statistics2}, panel a), as expected since we did not apply any 
k-correction to these samples. Both AXN and AXU samples show consistently 
the same trend (with slightly larger dispersions for the last bin, probably 
because this bin
is more subject to uncertainties due to the smaller numbers of AXU and AXN 
clusters there). This makes us confident that the samples are homogeneous at 
a first approximation and the techniques are adequate for the 
present study.

Other significant trends are the ones 
shown by the red sequence slope, which becomes flatter with decreasing redshift
(panel b of Figure \ref{statistics2}) and intercept, which consistently 
increases with redshift (panel c of Figure \ref{statistics2}). 
Some previous works \citep{Sta95, Sta98, San09, Mei09} have suggested, 
based on high redshift cluster samples, that there seems to be no 
change in the slope of the CMR from $z\,=\,$1.3 to $z\,=\,0$. However, 
other authors have already found a similar trend at a similar redshift 
range as the one considered in 
the present work. \cite{lopez} studied a sample of 57 Abell 
clusters of galaxies between 0.02 $\le\,z\,\le\,$0.18, selected by their 
X-ray emission. They show that their results are very well fitted by 
the models described by \cite{kodama97}, who concluded that the slope 
of the CMR is dominantly driven by a metallicity effect (instead of an 
age effect), taking its origin at early times probably from galactic 
wind feedback. The same conclusion was attained by \cite{Gladders}, who 
studied a sample of clusters, with data from Hubble Space Telescope, 
reaching higher redshifts (up to 0.75).

More than confirming a change in the slope of the CMR with redshift, 
for both AXUs and AXNs, panel b of Figure \ref{statistics2} also reveals that the 
slopes of the two samples are intrinsically (and significantly) different. 
The CMR slopes for the AXUs are systematically flatter than the CMR slopes 
of the AXNs. The mean difference is about 0.014 (69\%).

Our results on the intrinsic scatter of the CMR 
show an increase with redshift (Figure \ref{statistics2} panel d). 
This trend was also 
detected by \citet{vandokkum02} who found that the scatter is larger 
at much higher redshifts (0.4$\,<\,z\,<\,$0.7), with a tendency to decrease 
after that (0.7$\,<\,z\,<\,$1.3). Again, both of our 
samples have a similar behavior, but with a noticeable zero point difference, 
in the sense that the AXU sample presents higher scatter by 0.043 (28\%). 
Even considering that part of this effect could be produced by the increase
in the photometric errors with redshift, about 15\% along the studied redshift range, some
significant difference (more than 20\%) remains.

The fact that AXUs have probably less dense ICM may be related to
their dynamical status, or, being more specific, 
to their assembling history, which should be expected since in an 
hierarchical collapse the gravitational compression of the gas may
be slower than it would occur in a monolithic one.
The point here is that this different dynamical status seems to be
reflected in the 
evolutionary stage presented by the population of galaxies in clusters 
with ``normal'' (closer to relaxed) ICM or clusters with distinct 
intra-cluster characteristics.
As far as we know there are no previous references to this behavior. 

For estimating the significance of our results we applied the Cram\'er 
test \citep{baringhaus04}. For two distinct populations, this test estimates,
in our case, the probability that de distribution of slope, intrinsic scatter and
median colours differ between the samples (or more specifically we want to obtain the
probability of rejecting the hypothesis that the values come from the same parent 
population, in each case). 

In Table \ref{tabla4} we show the results for the Cram\'er test for the 
comparisons made in this work. We get a 92\% probability that the distributions 
of slope, for the AXN and AXU samples, are different, 78\% for the intercept, 
99\% for intrinsic scatter and 28\% for the median colour. 
From these results we can conclude that for the slope and intrinsic scatter we get
enough signal for a marginally significant difference between the two samples.

\begin{table}\small
\begin{center}
\begin{tabular}{lcccc}
\hline \hline 
\multicolumn{5}{c}{Table \ref{tabla4}. Results of Cram\'er test}\\ \hline
Sample & Slope & Intercept & Scatter & Colour \\
\hline
AXU $\times$ AXN & 92\% & 78\% & 99\% & 28\% \\
\hline \hline
\end{tabular}
\caption{{\small Results obtained from applying the Cram\'er test to our AXU
sample versus the AXN sample. The null hypothesis is that 
the compared samples come from the same parent population.}}
\label{tabla4}
\end{center}
\end{table}

\begin{figure}
\includegraphics[width=\linewidth]{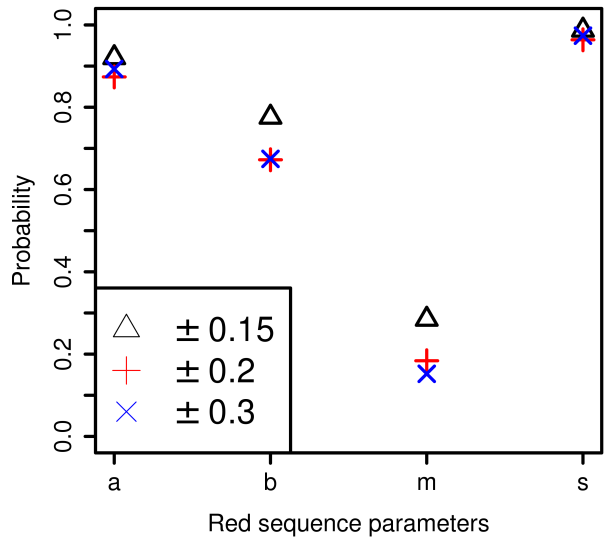}
\caption{Results of the Cram\'er probability test for different colour range 
box sizes on fitting the red sequence. Slope (a), intercept (b), median colour (m) and intrinsic scatter (s).}
\label{prob}
\end{figure}

Finally, we made an additional test to evaluate if our results were
dependent on the size of the selected box used to define the red sequence.
In Figure \ref{prob} we show the resulting Cram\'er probabilities for 
the difference between the AXU and AXN samples analyzed for 
distinct box size selections (in colour limits). One can see that the 
results do not change when we decrease the limits to 0.2 and 0.15. 

\section{Discussion}
\label{discussion}

One usual interpretation of the slope of the CMR, as we mentioned before, 
relates to the redder colours of the most luminous galaxies due to their 
higher metallicities, directly associated to the mass-metallicity relation 
\citep[e.g.][]{kodama97,kauffmann98}. In this sense, a galaxy population 
presenting a CMR with steeper slopes should have, naively speaking, 
their most massive galaxies overabundant in metals (or their less massive 
galaxies being ``submetallic''). In our samples, the steeper slopes of the 
red sequence on the CMRs are presented by the clusters with ``normal'' ICM, 
suggesting that the brightest galaxies in AXUs are less abundant in metals
(or the faintest are more metallic). 
This is difficult to reconcile with the scenario where the metallicity 
of the ICM comes from the feedback of the member galaxies. 
In the assumption that the AXU clusters are still galaxy systems 
in the process of formation, their underluminosity suggest that their ICM 
did not reach virial equilibrium yet and, thus, they are probably also 
underdense (in gravitational systems, collapse makes the gas hotter 
except only for very high opacities). Since the ram pressure, as an 
important hydrodynamical effect that removes gas (and metals) from the 
galaxies, depends on gas density, an underdense ICM should produce less 
ram pressure. Thus, the galaxies in this environment should be more 
metallic, instead, specially the most massive ones.
The correct scenario, thus, is probably much more complicated.

On the other hand, we can consider that the denser ICM (supposedly 
expected in a collapse closer to monolithic) may also produce 
more perturbation in the interestellar medium of the galaxies (by the same 
hydrodynamical effects) and will possibly stimulate the star formation in 
an initial phase (before it is sufficiently high to drag the cold gas and 
quench star formation). With more star formation, the galaxies produce 
more metals. In this sense, the metallicities of the galaxies are, on average,
affected by the properties of the ICM (and also by their star formation 
histories). This could be correct if the most affected galaxies in this 
scenario were the most massive ones. 
The enhancement of star formation is expected to happen mainly in early epochs 
of the cluster evolution, what would explain the parallel relations (the 
clusters with less dense ICM would have flatter CMRs along all their lives). 
This scenario is also in agreement with the results by \citet{Tan05} 
considering that the AXUs represent the lower density environments for which 
the CMR build-up is delayed.

Two questions remain open in this last scenario: the role of the feedback and 
typical metallicity of the ICM of underluminous clusters. We know that the 
metallicity of the AXNs is, on average, about 0.3 of the solar value, an already 
uncomfortably high value, but almost nothing is known about the metallicity of 
the AXU's ICM. 

Going into more detail, some recent models suggest that the fainter red-sequence (RS) 
galaxies have quenched their star formation only recently (and maybe suffered 
some events of wet mergers prior to that) and then settled in the RS 
\citep{Sma98, Tan05, Fab07, Mil12}.
Brighter RS galaxies, instead, are the prototype RS galaxies: they had the 
bulk of their stellar content already in place at $z >$ 1 and only evolved 
by major or minor dry mergers (of other RS galaxies) after that. 
According to \citet{jimenez11}, the effect of dry mergers is to increase the 
mass (and luminosity) of the galaxies without altering their metallicity 
\citep[see, also,][]{Bower3, Skelton}. These authors suggest that the bright 
end of the CMR must have a flat slope. In this scenario, AXUs may 
have flattened their RS by a higher rate of dry mergers. The question that 
remains is how to relate these rates with the shallower ICM.

\section{Conclusions}
\label{conclusion}
We compared the properties of the colour-magnitude relation, $(g-r) \times r$, 
of red galaxies in two samples: one with 56 X-ray underluminous clusters 
\citep[from][]{p1}, and the other with 50 ``normal'' X-ray emitting clusters 
\citep[from][]{p2}, in 5 bins of redshift. We have found that:

\begin{itemize}
\item[-] Our results confirm previous findings that the slope of the CMR 
decreases with increasing redshift. Both samples show the same trend, but 
the slopes of AXUs are always flatter than the slopes of AXNs, by a difference 
of about 69\% (0.014) along the surveyed redshift range.

\item[-] The intrinsic scatter of the CMR was found to grow with redshift, 
for both samples. Again, there is a zero point difference between 
the relations (intrinsic scatter $\times$ redshift) found for the AXUs and 
the AXNs, in the sense that the intrinsic scatter of the AXUs is systematically 
larger by more than 28\% (0.043).

\end{itemize}

Our data allow an interpretation such as that galaxies evolve in a distinct 
way in different environments concerning the properties of the ICM, which may be a consequence of the different dynamical status of the 
systems, being subject to different levels of star formation stimulation and, 
consequently, having different levels of metallicities, specially 
the more massive galaxies. 
In an initial phase of structure formation, a denser ICM would be expected
to enhance star formation in more massive galaxies, by exerting more hydrodynamical 
pressure. This scenario would lead the brightest galaxies in the AXN clusters
to be more metallic than their counterparts of the AXU clusters and, 
consequently, generating distinct slopes in their CMRs. If the X-ray 
underluminous clusters are, as previously proposed, less dense galaxy systems, 
this scenario is in accordance with the one proposed by \citet{Tan05}, in 
which the CMR build-up for less dense clusters is delayed. 
The larger intrinsic scatter for AXUs is also in agreement with this scenario: 
a longer time for producing the CMR implies also a larger dispersion in 
formation ages. A higher rate of dry mergers in X-ray underluminous clusters 
would also possibly produce the different slopes, but it is 
difficult to explain such an excess of interaction.

\section*{Acknowledgments}
JJTA thanks CONACyT for both a PhD and a beca mixta scholarships, and IAG 
USP staff for the hospitality during his academic visit. 
We thank the Brazilian agencies FAPESP and CNPq for their support to this 
work. We specially thank the anonymous referee for the very valuable comments 
and the quick replies. Funding for the SDSS and SDSS-II has been provided by 
the Alfred P. Sloan Foundation, the Participating Institutions, the National 
Science Foundation, the U.S. Department of Energy, the National Aeronautics 
and Space Administration, the Japanese Monbukagakusho, the Max Planck Society, 
and the Higher Education Funding Council for England. The SDSS Web Site is 
http://www.sdss.org/. The SDSS is managed by the Astrophysical Research 
Consortium for the Participating Institutions. This research has made use 
of the VizieR catalogue access tool, CDS, Strasbourg, France.

\thanks{This file has been amended to highlight the proper use of \LaTeXe code with the class file.}

\end{document}